\documentclass[12pt]{iopart}
\usepackage{graphicx}

\usepackage{iopams}  
\begin{document}

\title{Development behavior of liquid plasma produced by YAG laser}

\author{Jun Yamada$^*$ and Norio Tsuda}

\address{Dept. Electronics, Aichi Institute of Technology,\par
 1247 Yakusa-cho, Yachigusa, Toyota, 4700392, JAPAN\par
\textit{* E-mail:j-yamada@aitech.ac.jp}}

\begin{abstract}

The laser induced plasma in liquid hasn't been studied enough. In liquid, the laser induced plasma may be able to resolve the hazardous material called the environment material. Then, the plasma produced in liquid by the laser light is studied and the plasma development is observed by a streak camera. The ultra pure water or the ultra pure water with a melted NaCl is used as a test liquid. The liquid plasma is produced by the fundamental wave of YAG laser. When NaCl concentration is varied, the plasma development behavior is obserbed by streak camera. The liquid plasma develops backward. The plasma is produced from many seeds and It consists of a group of plasmas. However, the liquid plasma produced by second harmonic wave of YAG laser develops as a single plasma. The development mechanism is investigated from the growth rate of backward plasma. The backward plasma develops by breakdown wave and radiation supported shock wave.
\end{abstract}



\section{Introduction}

\hspace{10mm}When the laser light is focused at the solid or gas, a hot and dense plasma can be easily produced. Many studies of laser induced plasma with solid or gas targets have been down.\cite{1,2,3,4} However, the laser induced plasma in liquid hasn't been studied enough and the plasma development mechanisms have not almost been investigated. In liquid, the laser induced plasma may be able to resolve the hazardous material called the environment material. Then, the basic research of the plasma produced in liquid by the laser light is done and the development behavior of liquid plasma is observed by a streak camera.

\section{Experimental arrangement}

\hspace{10mm}The experiment arrangement to measure the plasma development behavior is shown in Fig. 1. The maximum laser power of YAG laser is 340 $mJ$ with a wavelength of 1064 $nm$ and a pulse half width of 15 $ns$. Moreover, the YAG laser is able to drive the second harmonic oscillation with a power of 180 $mJ$, a wavelength of 532 $nm$ and a pulse half width of 15 $ns$. The ultra pure water or the ultra pure water with a melted NaCl is used as a test liquid. The YAG laser light is focused in liquid from the out side of the chamber using the lens of the focal length 70 $mm$. The diameter of focal spot is 93 $\mu m$ for the fundamental wave. On the other hand, it is 54 $\mu m$ for the second harmonic wave. The laser power is controlled using the optical filter. The plasma development is observed using the streak camera. 

\section{Development behavior}
\subsection{Streak images}

\hspace{10mm}The typical streak images of the liquid plasma are shown in Fig. 2. The YAG laser is operated at 1064 $nm$ or 532 $nm$. The plasma produced in liquid develops only backward because the plasma frequency is higher than the laser frequency and laser light is absorbed at only backward plasma surface. The plasma consists of a group of plasmas produced from many seeds, which is produced by the fundamental wave. On the other hand, the plasma produced by the second harmonic of YAG laser is produced at the focal spot and develops backward as single plasma. The reason why the development behavior is different has not been understood well yet. 
The development of backward plasma was divided in two parts that is part 1 and part 2. Part 1 is fast development velocity started from focal spot, and part 2 is slow development velocity following with part 1. The backward development is compared with the theory.

\subsection{Plasma number}

\hspace{10mm}The number of backward plasma produced by fundamental wave of YAG laser is shown in Fig 3. The number of backward plasma increases with increasing NaCl concentration. On the other hand, when the light intensity is exceeded a breakdown threshold, the number of plasma increases rapidly with increasing light intensity. The electrolytes in liquid as seed may initiate the plasma production. 

\section{backward plasma development velocity}
\subsection{Theoretical calculation}

\hspace{10mm}The backward plasma development produced in liquid is theoriticaliy calculated. In high pressure gas, the backward plasma develops by breakdown wave and radiation supported shock wave. We thought the plasma induced in liquid develops same mechanism in high-pressure gas. 

\[
v_{b}\propto n_{g}^{1/2}W^{1/2}tan^{-1}\alpha \qquad v_{r}=\left[2\left(\gamma -1\right)\frac{WK}{\pi r^{2}m_{g}n_{g}}\right]^{1/3}
\]

These expressions show the theoretical development velocity of backward plasma in high-pressure gas. Where $v_{b}$ shows development velocity of the breakdown wave, $v_{r}$ one of the radiation supported shock wave, $W$ the laser power and $n_{g}$ the atomic density. The expression of breakdown wave shows that the development velocity is proportional to the $1/2$ power of the density and the laser power, and the radiation supported shock wave is proportional to the $-1/3$ power of the density and the $1/3$ power of the laser power. 

\subsection{Experimental data}

\hspace{10mm}The NaCl concentration dependence on backward development velocity in part 1 is shown in Fig. 4. The development velocities by fundamental wave is proportional to the $1/2$ power. On the other hand, the development velocity by second harmonic wave is proportional to the $-1/3$ power of the density below 1 $\%$ and it is proportional to the $1/2$ power of the density over 1 $\%$. This result shows that the development mechanism of the plasma in liquid differs by the NaCl solution.
The backward plasma development in part 2 may be diffusion because it does not depend except the NaCl concentration dependence.

\section{conculusion}

\hspace{10mm}The laser induced plasma in liquid hasn't been studied enough and the plasma development mechanisms have not almost been investigated. In liquid, the laser induced plasma may be able to resolve the hazardous material called the environment material. Then, the basic research of the plasma produced in liquid by the laser light is done.

Development behavior of liquid plasma is observed by streak camera. The plasma produced by fundamental YAG laser light consists of a group of plasmas. The plasma produced by second harmonic wave develops as a single plasma. The liquid plasma develops by breakdown wave and radiation supported shock wave same as development mechanisum of high-pressure plasma. 

\begin{flushleft}
REFERENCE
\end{flushleft}

\begin{figure}[h]
 \begin{center} 
  \centering   \includegraphics[width = 0.25 \linewidth]{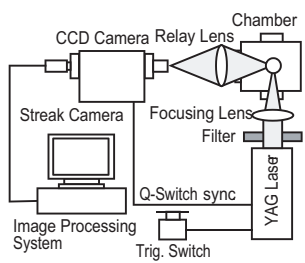} 
  \caption{Experimental arrangement} 
   \label{fig:1.eps} 
 \end{center}
\end{figure}

\begin{figure}[h]
 \begin{center} 
  \centering   \includegraphics[width = 0.7 \linewidth]{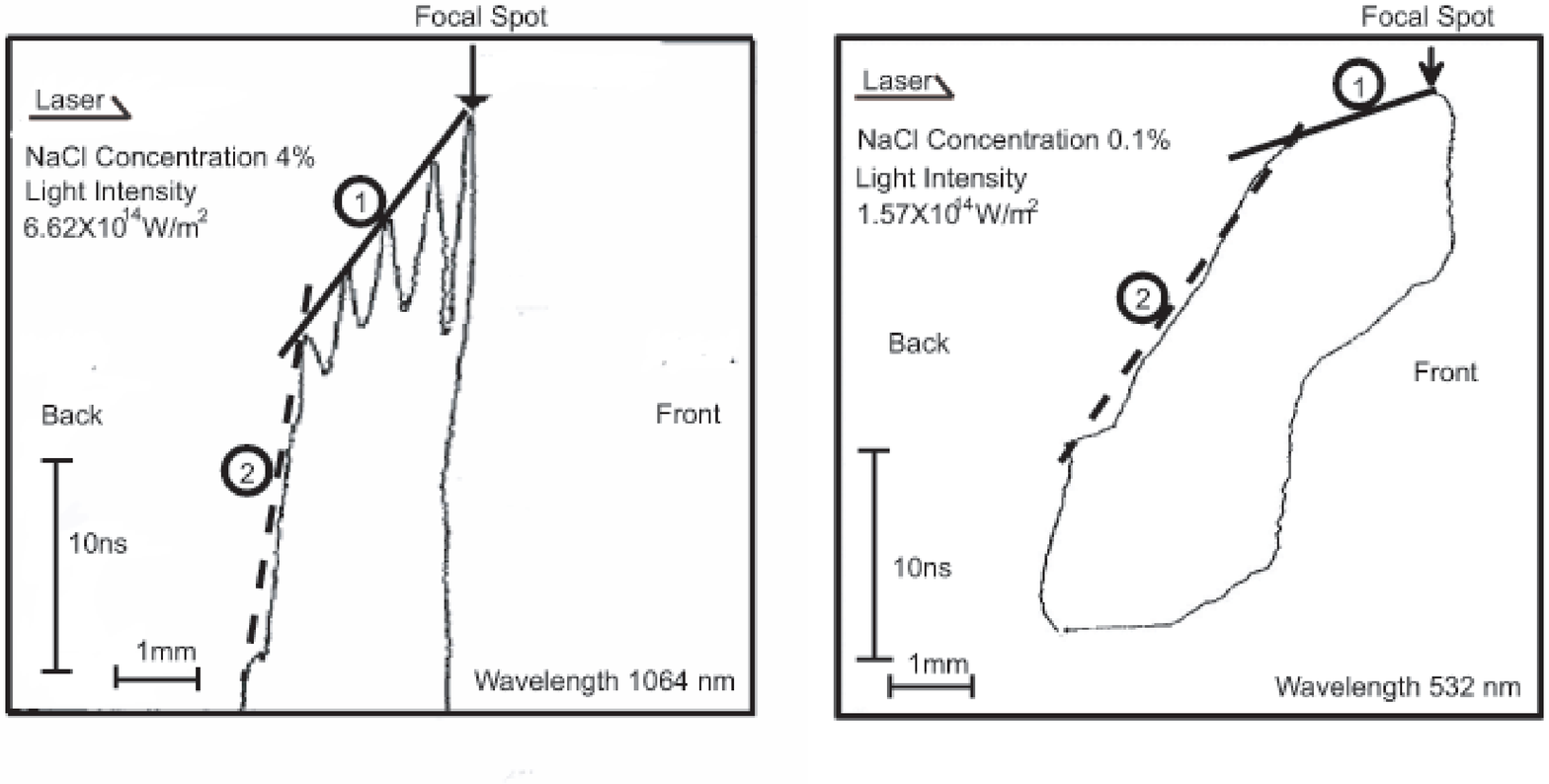} 
  \caption{Streak image} 
   \label{fig:2.eps} 
 \end{center}
\end{figure}

\begin{figure}[h]
 \begin{center} 
\includegraphics[width = 0.32 \linewidth]{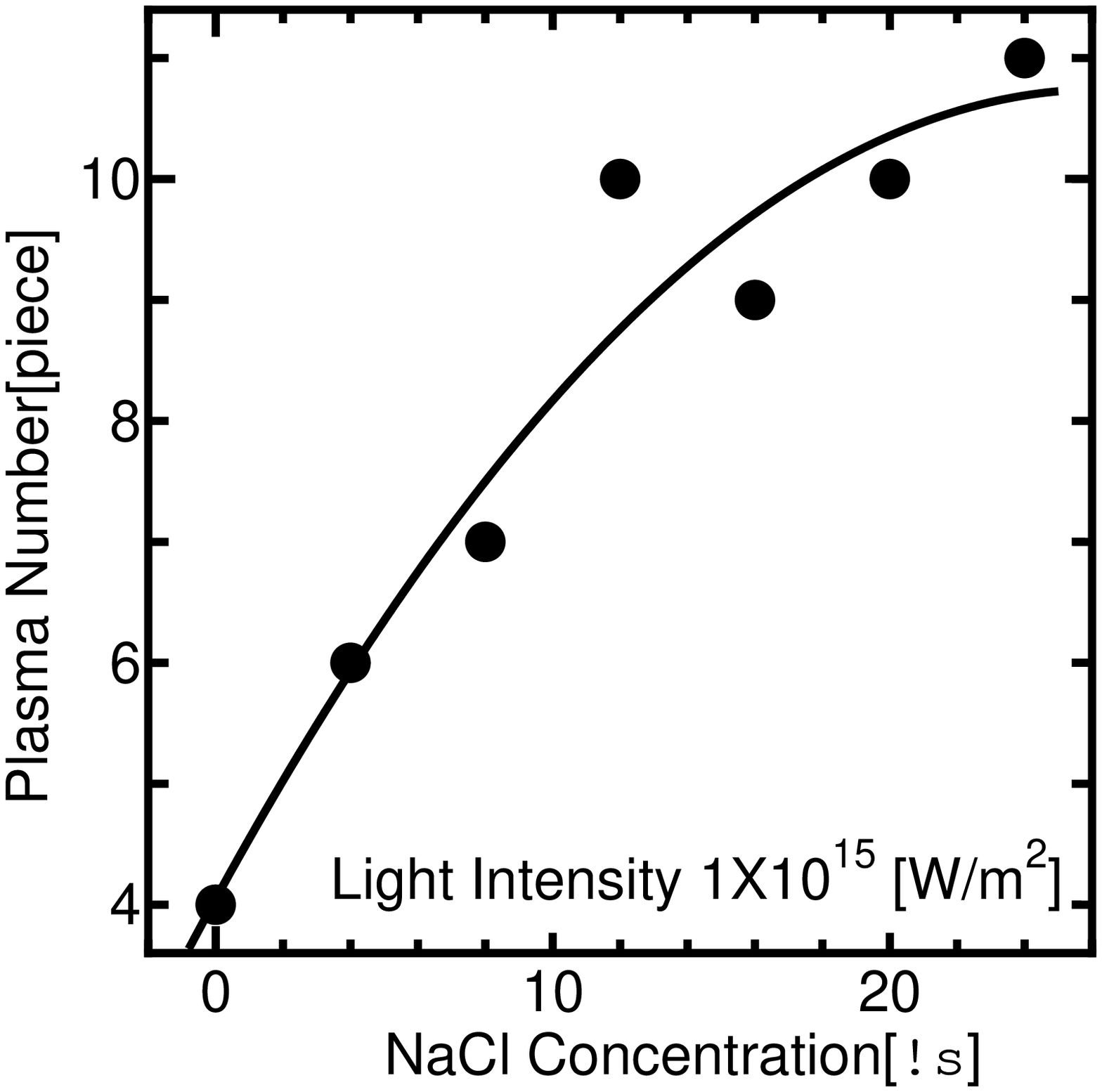}
\includegraphics[width = 0.4 \linewidth]{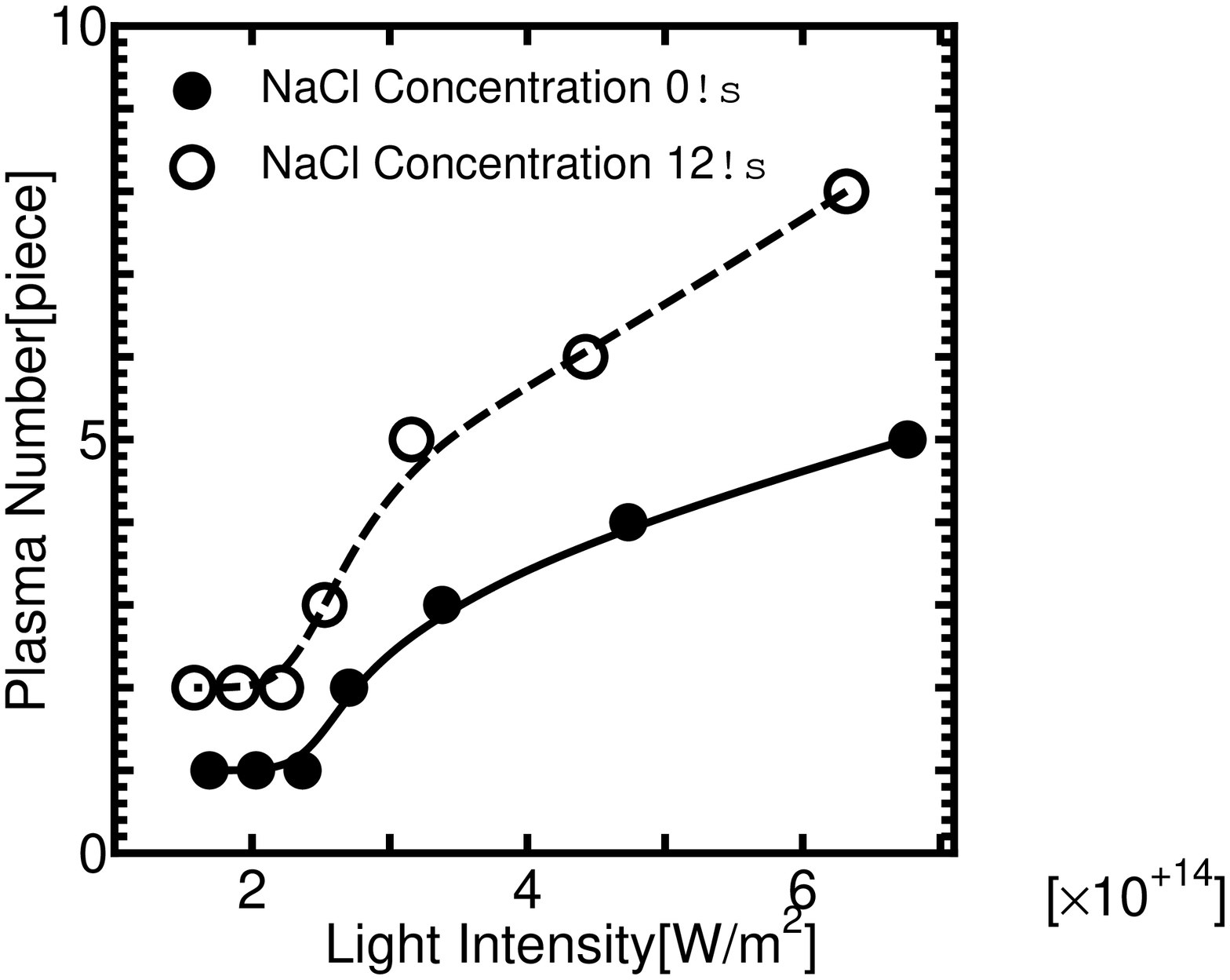}
  \caption{Plasma number} 
   \label{fig:3.eps} 
 \end{center}
\end{figure}

\begin{figure}[h]
 \begin{center} 
\includegraphics[width = 0.35 \linewidth]{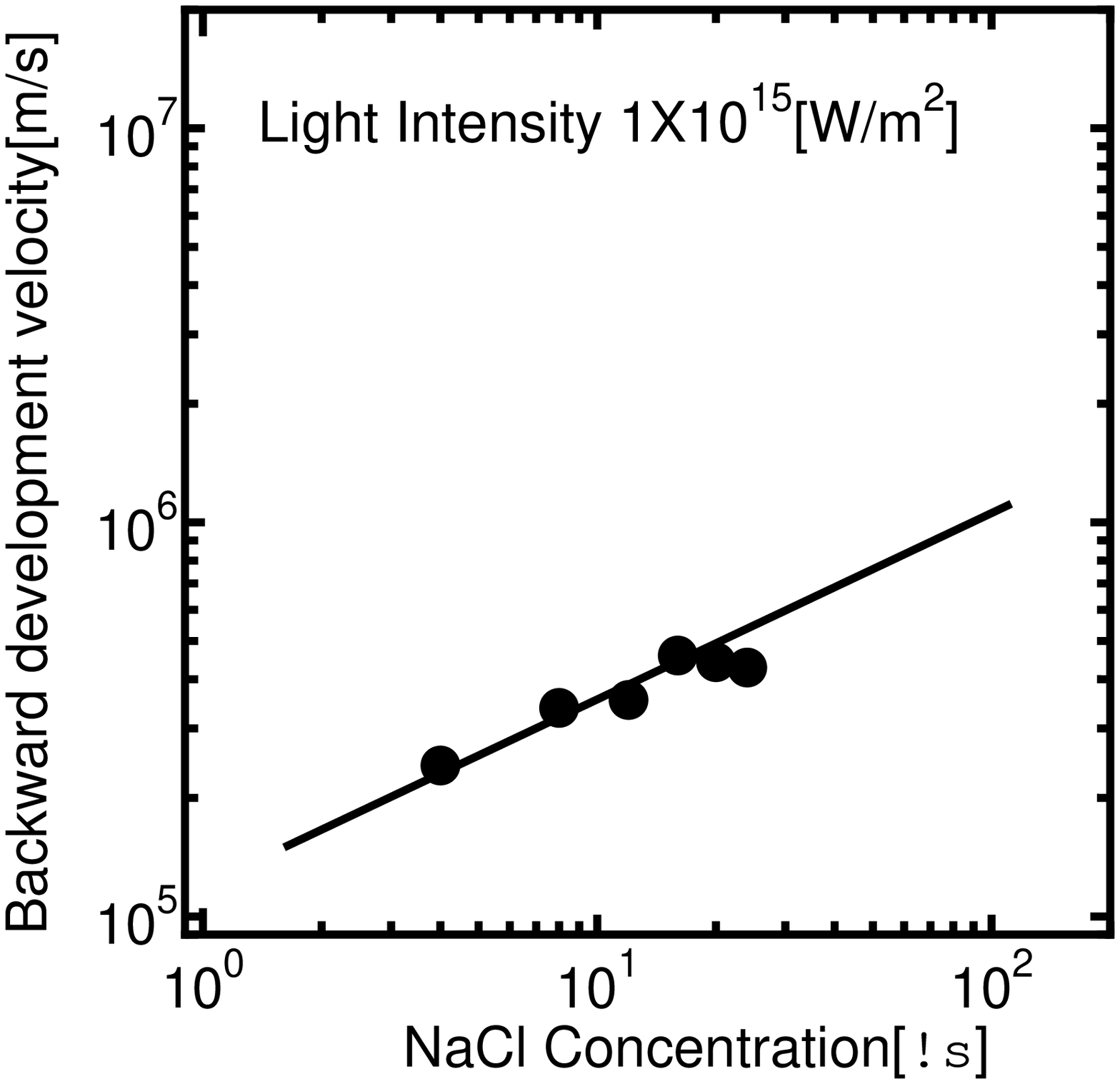}
\includegraphics[width = 0.35 \linewidth]{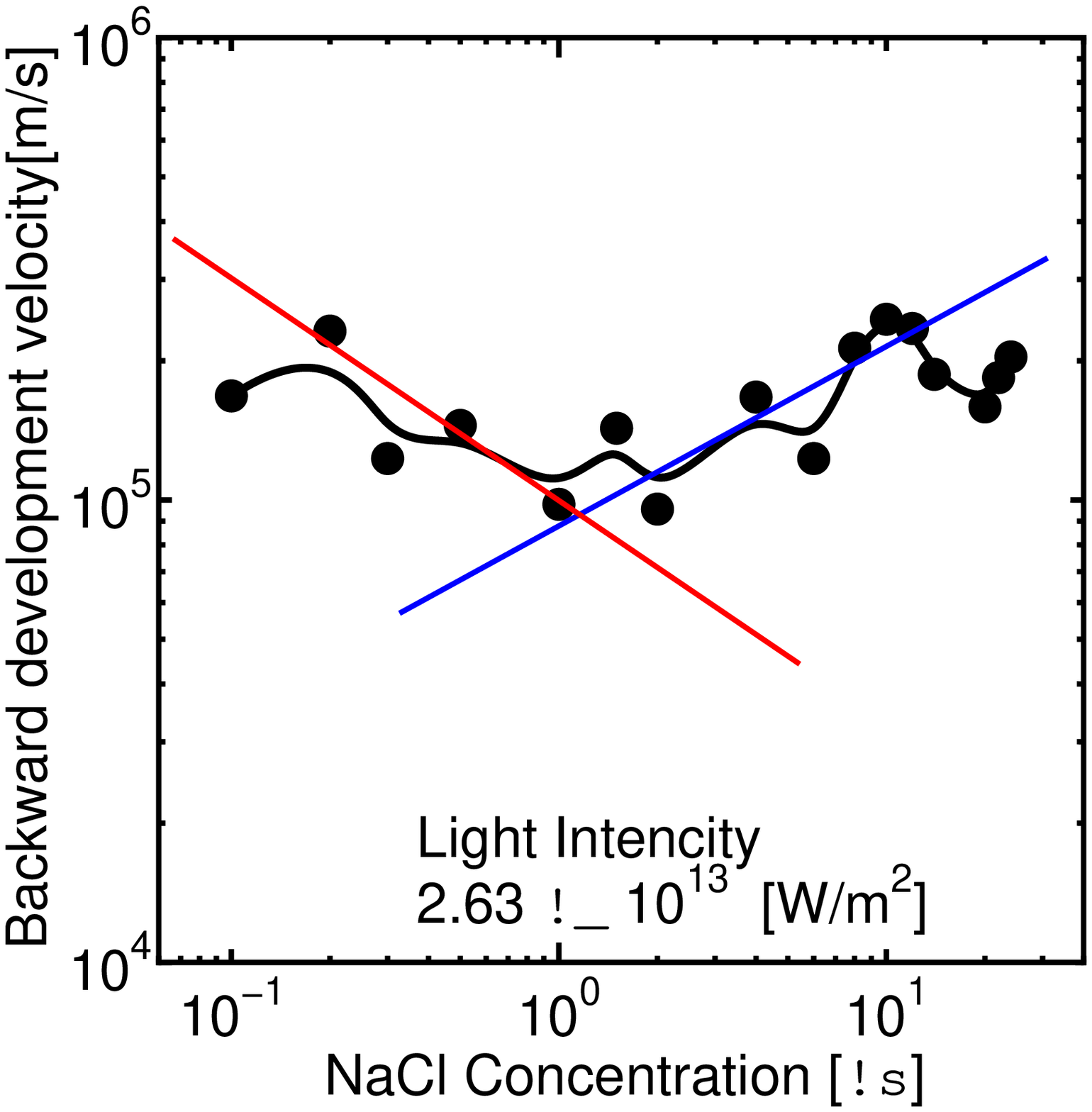}
  \caption{Backward plasma development velocity} 
   \label{fig:4.eps} 
 \end{center}
\end{figure}

\end{document}